\documentclass[runningheads]{llncs}
\usepackage{graphicx}
\usepackage{soul}
\usepackage[utf8]{inputenc}
\usepackage[table]{xcolor}
 \usepackage{cancel} 
\usepackage{float}
\usepackage{graphicx}
\usepackage{xcolor}
\usepackage[sort,nocompress]{cite}
\usepackage{todonotes}
\usepackage{verbatim}
\usepackage{subcaption}
\usepackage[normalem]{ulem}
\usepackage{array}
\usepackage{mathrsfs}
 \usepackage{multirow}
 \usepackage{blindtext}
\usepackage{hyperref}
\usepackage{graphicx}
\usepackage{amsmath}
\usepackage{mathtools}
\usepackage{mathrsfs}
 \usepackage{multirow}
 \usepackage{blindtext}
\usepackage{hyperref}
\usepackage{graphicx}
\usepackage{algorithm}
\usepackage{algpseudocode}

\usepackage{pgf}
\usepackage{tikz}
\usepackage{tikz-cd}
\tikzcdset{scale cd/.style={every label/.append style={scale=#1},
    cells={nodes={scale=#1}}}}

\usetikzlibrary{arrows,automata,positioning}
\newcommand{\LCP}{\mathsf{LCP}}
\newcommand{\lcp}{\mathop{\mathsf{lcp}}}
\newcommand{\initstate}{s_0}

\DeclarePairedDelimiter{\ceil}{\lceil}{\rceil}

\newcommand{\notaNC}[1]{\textcolor{blue}{#1 NC}}

\begin{document}
\title{Space-time Trade-offs for the LCP Array of Wheeler DFAs
}
%
%
\author{Nicola Cotumaccio \inst{1, 2} \orcidID{0000-0002-1402-5298} \and
Travis Gagie \inst{2}\orcidID{0000-0003-3689-327X} \and
Dominik Köppl \inst{3}\orcidID{0000-0002-8721-4444} \and
Nicola Prezza \inst{4}\orcidID{0000-0003-3553-4953}
}
\authorrunning{N. Cotumaccio et al.}
%
\institute{GSSI, Italy \email{nicola.cotumaccio@gssi.it} \and
Dalhousie University, Canada 
\email{nicola.cotumaccio@dal.ca, travis.gagie@dal.ca}\\ \and
University of Münster, Germany \email{dominik.koeppl@uni-muenster.de} \\ \and
University Ca’ Foscari, Venice, Italy \email{nicola.prezza@unive.it} \\}
\maketitle              
\begin{abstract}
Recently, Conte et al. generalized the longest-common prefix (LCP) array from strings to Wheeler DFAs, and they showed that it can be used to efficiently determine matching statistics on a Wheeler DFA [DCC 2023]. However, storing the LCP array requires $ O(n \log n) $ bits, $ n $ being the number of states, while the compact representation of Wheeler DFAs often requires much less space. In particular, the BOSS representation of a de Bruijn graph only requires a linear number of bits, if the size of alphabet is constant.

In this paper, we propose a sampling technique that allows to access an entry of the LCP array in logarithmic time by only storing a linear number of bits. We use our technique to provide a space-time trade-off to compute matching statistics on a Wheeler DFA. In addition, we show that by augmenting the BOSS representation of a $ k $-th order de Bruijn graph with a linear number of bits we can navigate the underlying variable-order de Bruijn graph in time logarithmic in $ k $, thus improving a previous bound by Boucher et al. which was linear in $ k $ [DCC 2015].

\end{abstract}

\section{Introduction}

In 1973, Weiner invented the \emph{suffix tree} of a string \cite{weiner1973}, a versatile data structure which allows to efficiently handle a variety of problems, including solving pattern matching queries, determining matching statistics, identifying combinatorial properties of the string and computing its Lempel-Ziv decomposition. However, the space consumption of a suffix tree can be too high for some applications (including bioinformatics), so over the past 30 years a number of \emph{compressed} data structures simulating the behavior of a suffix tree have been designed, thus leading to compressed suffix trees \cite{sadakane2007}. In many applications, one does not need the full functionality of a suffix tree, so it may be sufficient to store only some of these data structures. Among the most popular data structures, we have the suffix array \cite{manber1993}, the longest common prefix (LCP) array \cite{manber1993}, the Burrows-Wheeler transform (BWT) \cite{burrows1994} and the FM-index \cite{ferragina2000}.

In the past 20 years, the ideas behind the suffix array, the BWT and the FM-index have been generalized to trees \cite{ferragina2005, ferragina2009}, de Bruijn graphs \cite{BOSS}, Wheeler graphs \cite{gagie2017, alanko2020} and arbitrary graphs and 
automata \cite{cotumaccio2021, cotumaccio2022}. Broadly speaking, Wheeler graphs concisely capture the intuition behind these data structures in a graph setting; thus, they can be regarded as a benchmark for extending suffix tree functionality to graphs. In particular, the LCP array of a string remarkably extends the functionality of the suffix array, and a recent paper \cite{conte2023} shows that the LCP array can also be generalized to Wheeler DFAs, which represents a remarkable step toward fully simulating suffix-tree functionality in a graph setting. However, the solution in \cite{conte2023} is not space efficient: storing the LCP array of a Wheeler DFA requires $ O(n \log n) $ bits, $ n $ being the number of states. If the size $ \sigma $ of the alphabet is small, this space can be considerably larger than the space required to store the Wheeler DFA itself. As we will see, if $ \sigma \log \sigma = o(\log n) $
, then the space required to store the Wheeler DFA is $ o (n \log n) $, 
and if $ \sigma = O(1) $, then the space required to store the Wheeler DFA is $ O(n) $. The latter case is especially relevant in practice, because de Bruijn graphs are the prototypes of Wheeler graphs, and in bioinformatics de Bruijn graphs are defined over the constant-size alphabet $ \Sigma = \{A, C, G, T \} $.

In this paper, we show that we can \emph{sample} entries of the LCP array in such a way that, by storing only a linear number of additional bits on top of the Wheeler graph, we can compute each entry of the LCP array in logarithmic time, thus providing a space-time trade-off. More precisely:

\begin{theorem}\label{theor:timespacetradeoff}
    We can augment the compact representation of a Wheeler DFA $ \mathcal{A} $ with $ O(n) $ bits ($ O(n \log \log \sigma) $ bits, respectively), where $ n $ is the number of states and $ \sigma $ is the size of the alphabet, in such a way that we can compute each entry of the LCP array of $ \mathcal{A} $ in $ O(\log n \log \log \sigma) $ time ($ O (\log n) $ time, respectively).
\end{theorem}


We present two applications of our result: computing matching statistics on Wheeler DFAs and navigating varriable-order de Bruijn graphs.

\subsubsection*{Matching Statistics on Wheeler DFAs}

The problem of computing matching statistics on a Wheeler DFA is defined as follows: given a pattern of length $ m $ and a Wheeler DFA with $ n $ states, determine the longest suffix of each prefix of $ m $ that occurs in the graph (that is, that can be read by following some edges on the graph and concatenating the labels). This problem is a natural generalization of the problem of computing matching statistics on strings. Conte et al.~\cite{conte2023} proved the following result:
   \begin{theorem}\label{theor:oldmatching}
    We can augment the compact representation of a Wheeler DFA $ \mathcal{A} $ with $ O(n \log n) $ bits, where $ n $ is the number of states and $ \sigma $ is the size of the alphabet, in such a way that we can compute the matching statistics of a pattern of length $ m $ w.r.t to the Wheeler DFA in $ O(m \log n) $ time.
\end{theorem} 

We will show that if we only want to use linear space, then we can use Theorem \ref{theor:timespacetradeoff} to obtain the following trade-off.

\begin{theorem}\label{theor:newmatching}
    We can augment the compact representation of a Wheeler DFA $ \mathcal{A} $ with $ O(n \log \log \sigma) $ bits, where $ n $ is the number of states and $ \sigma $ is the size of the alphabet, in such a way that we can compute the matching statistics of a pattern of length $ m $ w.r.t to the Wheeler DFA in $ O(m \log^2 n) $ time.
\end{theorem}

\subsubsection*{Variable-order de Bruijn Graphs} 
Wheeler graphs are a generalization of de Bruijn graphs; in particular, the compact representation of a Wheeler graph is a generalization of the BOSS representation of a de Bruijn graph \cite{BOSS}, and our results on the LCP array also apply to a de Bruijn graph. 
Many assemblers \cite{bankevich2012, peng2010, li2009, simpson2009} consider all $ k $-mers occurring in a set of reads and build a $ k $-th order de Bruijn graph (on the alphabet $ \Sigma = \{A, C, G, T \} $) to perform Eulerian sequence assembly \cite{Idury1995, pavel2001}. However, the choice of the parameter $ k $ impacts the assembly quality, so some assemblers try several choices for $ k $ \cite{bankevich2012, peng2010}, which slows down the process because several de Bruijn graphs need to be built. In \cite{boucher2015} it was shown that the $ k $-order de Bruijn graph of $ \mathcal{S} $ can be used to \emph{implicitly} store the $ k' $-th order de Bruijn graph of $ \mathcal{S} $ for \emph{every} $ k' \le k $, thus leading to a \emph{variable-order de Bruijn graph}. The challenge is to navigate this implicit representation (that is, how to follow edges in a forward or backward fashion). In \cite{boucher2015}, it was shown that the navigation is possible by storing or by simulating an array $ \overline{\LCP}_G $ which can be seen as a simplification of the LCP array of the Wheeler graph $ G $. More precisely, we have the following result (see \cite{boucher2015}; we assume $ \sigma = O(1) $).

\begin{theorem}\label{theor:oldvaribale}
    \begin{enumerate}
        \item We can augment the BOSS representation of a $ k $-th order de Bruijn graph with $ O(n \log k ) $ bits, where $ n $ is the number of nodes, so that the underlying variable-order de Bruijn graph can be navigated in $ O(\log k ) $ time per visited node.
        \item We can augment the BOSS representation of a $ k $-th order de Bruijn graph with $ O(n) $ bits, where $ n $ is the number of nodes, so that the underlying variable-order de Bruijn graph can be navigated in $ O(k \log n ) $ time per visited node.
    \end{enumerate}
\end{theorem}

Essentially, the first solution in Theorem \ref{theor:oldvaribale} explicitly stores $ \overline{\LCP}_G $, while the second solution in Theorem \ref{theor:oldvaribale} computes the entries of $ \overline{\LCP}_G $ by exploiting the BOSS representation. In general, a big $ k $ (close to the size of the reads) allows to retrieve the expressive power on an overlap graph \cite{dazdomnguez2019}, so in Theorem \ref{theor:oldvaribale} we cannot assume that $ k $ is small. On the one hand, the \emph{space} required for the first solution can be too large, because a de Bruijn graph can be stored by using only $ O(n) $ bits. On the other hand, the \emph{time} bound in the second solution increases substantially. We can now improve the second solution by providing a data structure that achieves the best of both worlds. As we did in Theorem \ref{theor:timespacetradeoff}, we can conveniently sample some entries of $ \overline{\LCP}_G $. We will prove the following result.

\begin{theorem}\label{theor:newvariable}
   We can augment the BOSS representation of a $ k $-th order de Bruijn graph with $ O(n) $ bits, where $ n $ is the number of nodes, so that the underlying variable-order de Bruijn graph can be navigated in $ O(\log k \log n ) $ time per visited node. 
\end{theorem}

\section{Definitions}\label{sec:definitions}

\subsubsection*{Sets and Relations} Let $ V $ be a set. A total order on $ V $ is a binary relation $ \le $ which is reflexive, antisymmetric and transitive. We say that $ U $ is a $ \le $-interval (or simply an interval) if for all $ v_1, v_2, v_3 \in V $, if $ v_1, v_3 \in U $ and $ v_1 < v_2 < v_3 $, then $ v_2 \in U $. If $ u, v \in V $, with $ u \le v $, we denote by $ [u, v] $ the smallest interval containing $ u $ and $ v $, that is $ [u, v] = \{z \in V \;|\; u \le z \le v $ \}. In particular, if $ V $ is the set of integers, then we assume that $ \le $ is the standard total order, hence $ [u, v] = \{u, u + 1, \dots, v - 1, v \} $.

\subsubsection*{Strings} Let $ \Sigma $ be a finite alphabet, with $ \sigma = |\Sigma|$. Let $ \Sigma^* $ be the set of all finite strings on $ \Sigma $ and let $ \Sigma^\omega $ be the set of all (countably) infinite strings on $ \Sigma $.
If $ \alpha \in \Sigma^* $, then $ \alpha^R $ is the reverse string of $ \alpha $. If $ \alpha, \beta \in \Sigma^* \cup \Sigma^\omega $, we denote by $ \lcp (\alpha, \beta) $ the length of longest common prefix between $ \alpha $ and $ \beta $. In particular, if $ \alpha \in \Sigma^* $, then $ \lcp (\alpha, \beta) \le |\alpha| $ and if $ \alpha, \beta \in \Sigma^\omega $ with $ \alpha = \beta $, then $ \lcp (\alpha, \beta) = \infty $. 
Let $ \preceq $ be a fixed total order on $ \Sigma $. We extend the total order $ \preceq $ from $ \Sigma $ to $ \Sigma^* \cup \Sigma^\omega $ lexicographically.

\subsubsection*{DFAs} Throughout the paper, let $ \mathcal{A} = (Q, E, \initstate, F) $ be a deterministic finite automaton (DFA), where $ Q $ is the set of states, $ E \subseteq Q \times Q \times \Sigma $ is the set of labeled edges, $ \initstate \in Q $ is the initial state and $ F \subseteq Q $ is the set of final states. The alphabet $ \Sigma $ is \emph{effective}, that is, every $ c \in \Sigma $ labels some edge. Since $ \mathcal{A} $ is deterministic, for every $ u \in Q $ and for every $ a \in \Sigma $ there exists at most one edge labeled $ a $ leaving $ u $. Following \cite{alanko2020}, we assume that (i) $ \initstate $ has no incoming edges, (ii) every state is reachable from the initial state and (iii) all edges entering the same state have the same label (\emph{input-consistency}). For every $ u \in Q \setminus \{\initstate\} $, let $ \lambda (u) \in \Sigma $ be the label of all edges entering $ u $. 
We define $ \lambda (\initstate) = \# $, where $ \# \not \in \Sigma $ is a special character such that $ \# \prec a $ for every $ a \in \Sigma $ (the character $ \# $ plays the same role as the termination character $ \$ $ in suffix arrays, suffix trees and Burrows-Wheeler transforms). As a consequence, an edge $ (u', u, a) $ can be simply written as $ (u', u) $, because it must be $ a = \lambda (u) $.

\subsubsection*{Compact Data Structures} Let $ A $ be an array of length $ n $ containing elements from a finite totally-ordered set. A \emph{range minimum query}  on $ A $ is defined as follows: given $ 1 \le i \le j \le n $, return one of the indices $k$ with $ 1 \le k \le n $ such that (i) $ i \le k \le j $ and $ A[k] = \min\{A[i], A[i + 1], \dots, A[j - 1], A[j] \} $. We write $ k = RMQ_A(i, j) $. Then, there exists a data structure of $ 2n + o(n) $ such that in $ O(1) $ time we can compute $ RMQ_A(i, j) $ for every $ 1 \le i \le n $, \emph{without the need to access $ A $} \cite{fischer2010, fischer2011}. This result is essentially optimal, because every data structure solving range minimum queries on $ A $ requires at least $ 2n - \Theta ( \log n) $ bits \cite{fischer2011, liu2020}.

Let $ A  $ be a bitvector of length $ n $. Let $ rank(A, i) = |\{j\in \{1, 2, \dots, i - 1, i \} \;|\; A[j]=1 \}|$ be the number of $ 1 $'s among the first $ i $ bits of $ A $.
Then, there exists a data structure of $ n + o (n) $ bits such that in $ O (1) $ time we can compute $ rank(A, i) $ for $ 1 \le i \le n $ \cite{navarro2016}.

\section{Wheeler DFAs}

Let us recall the definition of Wheeler DFA \cite{conte2023}.

\begin{definition}\label{def:wheeler}
Let $ \mathcal{A} = (Q, E, \initstate, F) $ be a DFA. A \emph{Wheeler order} on $ \mathcal{A} $ is a total order $ \le $ on Q such that $ \initstate \le u $ for every $ u \in Q $ and:
\begin{enumerate}
    \item (Axiom 1) If $ u,v \in Q $ and $ u < v $, then $ \lambda(u) \preceq \lambda (v) $.
    \item (Axiom 2) If $ (u', u), (v', v) \in E $, $ \lambda (u) = \lambda (v)  $ and $ u < v $, then $ u' < v' $.
\end{enumerate}
A DFA $\mathcal{A}$ is \emph{Wheeler} if it admits a Wheeler order.
\end{definition}

Every DFA admits at most one Wheeler order \cite{alanko2020}, so the total order $ \le $ in Definition \ref{def:wheeler} is \emph{the} Wheeler order on $ \mathcal{A} $. In the following, we fix a Wheeler DFA $ \mathcal{A} = (Q, E, \initstate, F) $, with $ n = |Q| $ and $ e = |E| $, and we write $ Q = \{u_1, \dots, u_n \} $, with $ u_1 < u_2 < \dots < u_n $ in the Wheeler order. In particular, $ u_1 = \initstate $. Following \cite{conte2023}, we assume that $ \initstate $ has a self-loop labeled $ \# $, which is consistent with Axiom 1, because $ \# \prec a $ for every $ a \in \Sigma $). This implies that every state has at least one incoming edge, so for every state $ u_i $ there exists at least one infinite string $ \alpha \in \Sigma^\omega $ that can be read starting from $ u_i $ and following edges in a backward fashion. We denote by $ I_{u_i} $ the nonempty set of all such strings. Formally:

\begin{definition}
Let $ 1 \le i \le n $. Define:
\begin{equation*}
\begin{split}
    I_{u_i} = \{\alpha \in \Sigma^\omega \; | \; \text{there exist integers $ f_1, f_2 , \dots $ in $ [1, n] $ such that (i) $ f_1 = i $,} \\
    \text{(ii) $ (u_{f_{k + 1}}, u_{f_k}) \in E $ for every $ k \ge 1 $ and (iii) $ \alpha = \lambda (u_{f_1}) \lambda (u_{f_2}) \dots $} \}.
\end{split}    
\end{equation*}
\end{definition}

For every $ 1 \le i \le n $, let $ p_{\min} (i) $ be the smallest $ 1 \le i' \le n $ such that $ (u_{i'}, u_i) \in E $ and let $ p_{\max} (i) $ be the largest $ 1 \le i'' \le n $ such that $ (u_{i''}, u_i) \in E $. Both $ p_{\min} (i) $ and $ p_{\max} (i) $ are well-defined because every state has at least one incoming edge. For every $ 1 \le i \le n $, define $ p^1_{\min} (i) = p_{\min} (i) $ and recursively, for $ j \ge 2 $, let $ p_{\min}^j(i) = p_{\min}(p_{\min}^{j - 1}(i)) $. Then, $ \lambda (u_i) \lambda (p_{\min} (i)) \lambda (p_{\min}^2(i)) \lambda (p_{\min}^3(i)) \dots  $ is the lexicographically \emph{smallest} string in $ I_{u_i} $, which we denote by $ \min_i $ \cite{conte2023}. Analogously, one can define the lexicographically \emph{largest} string in $ I_{u_i} $ by using $p_{\max}$. Moreover, in \cite{conte2023} it was shown that:
\begin{equation*}
    \mathrm{min}_1 \preceq \mathrm{max}_1 \preceq \mathrm{min}_2 \preceq \mathrm{max}_2 \preceq \dots \preceq \mathrm{max}_{n - 1} \preceq \mathrm{min}_n \preceq \mathrm{max}_n.
\end{equation*}

Intuitively, the previous equation shows that the permutation of the set of all states of $ \mathcal{A} $ induced by the Wheeler order can be seen as a generalization of the permutation of positions induced by the prefix array of a string $ \alpha $ (or equivalently, the suffix array of the reverse string of $ \alpha^R $). Indeed, a string $ \alpha $ can also be seen as a DFA $ \mathcal{A}' = (Q', E', \initstate', F') $, where $ Q' = \{q'_0, q'_1 \dots, q'_{|\alpha|} \} $, $ \initstate' = q'_0 $, $ F' = \{q'_{|\alpha|}  \} $ (the set $ F $ plays no role here), $ \lambda (q'_i) $ is the $ i $-th character of $ \alpha $ for $ 1 \le i \le n $ and $ E' = \{(q'_{i - 1}, q'_i) \;|\; 1 \le i \le n \} $ (every state is reached by exactly one string so the minimum and the maximum string reaching each state are equal).

Let $ 1 \le r \le s \le n $ and let $ c \in \Sigma $. Let $ E_{r, s, c} $ be the set of all states that can be reached from a state in $ [r, s] $ by following edges labeled $ c $; formally, $ E_{r, s, c}= \{1 \le j \le n \;|\; \text{$ \lambda (u_j) = c $ and $ (u_i, u_j) \in E $ for some $ i \in [r, s] $ } \}$. 
Then, $ E_{r, s, c} $ is again an interval, that is, there exist $ 1 \le r' \le s' \le n $ such that $ E_{r, s, c} = [r', s'] $ \cite{gagie2017}. This property enables a compression mechanism that generalizes the Burrows-Wheeler transform \cite{burrows1994 } and the FM-index \cite{ferragina2000} to Wheeler DFAs. 
The Wheeler DFA $ \mathcal{A} $ can be stored by using only $ 2e + 4n + e \log \sigma + \sigma \log e $ bits (up to lower order terms), including $ n $ bits to mark the set $ F $ of final states and $ n $ bits to mark all $ 1 \le i \le n $ such that $ \lambda (u_i) \not = \lambda (u_{i - 1}) $, which allows us to retrieve each $ \lambda (u_i) $ in $ O(1) $ time by using a rank query \cite{gagie2017} (recall that $ n $ is the number of states and $ e $ is the number of edges). Since $ \mathcal{A} $ is a DFA, we have $ e \le n \sigma $, so the required space is $ O(n \sigma \log \sigma) $. If the alphabet is small --- that is, if $ \sigma \log \sigma = o(\log n) $
--- then the number of required bits is $ o (n \log n ) $; if $ \sigma = O(1) $, then the number of required bits is $ O(n) $. This compact representation supports efficient navigation of the graph and it allows to solve pattern matching queries. More precisely, by resorting to state-of-the art select queries \cite{navarro2016} in $ O (\log \log \sigma) $ time (i) for $ 1 \le i \le n $, we can compute $ p_{\min} (i) $ and $ p_{\max} (i) $ and (ii) given $ 1 \le r \le s \le n $ and $ c \in \Sigma $, we can compute $ [r', s'] = E_{r, s, c} $ \cite{gagie2017}. In particular, query (ii) is the so-called \emph{forward-search}, which generalizes the analogous mechanism of the FM-index, thus allowing to solve pattern matching queries on the graph.

The Wheeler order generalizes the notion of suffix array from strings to DFA. It is also possible to generalize LCP-arrays from strings to graph \cite{conte2023}.

\begin{definition}
    The \emph{LCP-array} of the Wheeler DFA $ \mathcal{A} $ is the array $ \LCP_\mathcal{A} = \LCP_\mathcal{A}[2, 2n] $ which contains the following $ 2n - 1 $ values in the following order: $ \lcp(\min_1, \max_1) $, $ \lcp(\max_1, \min_2) $, $ \lcp(\min_2, \max_2) $, $ \dots $, $ \lcp(\max_{n - 1}, \min_n) $, $ \lcp (\min_n, \max_n) $. In other words, $ \LCP [2i] = \lcp (\min_i, \max_i) $ for $ 1 \le i \le n $ and $ \LCP_\mathcal{A}[2i - 1] = \lcp(\max_{i - 1}, \min_i) $ for $ 2 \le i \le n $.
\end{definition}

It can be proved that for every $ 2 \le i \le n $, if $ \LCP_\mathcal{A}[i] $ is finite, then $ \LCP_\mathcal{A}[i] < 3n $ \cite{conte2023}. As a consequence, $ \LCP_\mathcal{A} $ can be stored by using $ O(n \log n) $ bits.

\section{A Space-time Trade-off for the LCP Array}

By storing an LCP array on top of the compact representation of a Wheeler graph, we have additional information that we can use to efficiently solve problems such as computing the matching statistics; however, we need to store $ O(n \log n) $ bits. As we have seen, $ O(n \log n) $ dominates the number of bits required to store $ \mathcal{A} $ itself, if the alphabet is small. In this section, we show that we can store a data structure of only $ O(n \log \log \sigma) $ bits which allows to compute every entry $ \LCP_\mathcal{A}[i] $ in $ O(\log n ) $ time, thus proving Theorem \ref{theor:timespacetradeoff}. This will be possible by sampling some entries of $ \LCP_\mathcal{A} $. The sampling mechanism is obtained by conveniently defining an auxiliary graph from the entries of the LCP array. We will immediately describe our technique, our sampling mechanism being general-purpose.

\begin{algorithm}[!t]
\scriptsize
\caption{Building $ V(h) $}\label{alg:V(h)construction}
\begin{algorithmic}
\State $ V(h) \gets \emptyset $
\State $ U \gets \emptyset $
\While{there exists $ v \in V $ such that (a) $ v(i) $ is defined for $ 0 \le i \le h - 1 $, (b) $ v(i) \not = v(j) $ for $ 0 \le j < i \le h - 1 $, (c) $ v(i) \not \in U $ for $ 0 \le i \le h - 1 $}
    \State Pick such a $ v $, add $ v(h - 1) $ to $ V(h) $ and add $ v(i) $ to $ U $ for every $ 0 \le i \le h - 1 $
\EndWhile
\end{algorithmic}
\end{algorithm}

\begin{algorithm}[!t]
\scriptsize
\caption{Input: $ h \in [2, 2n] $. Output: $ \LCP_\mathcal{A}[h] $.}\label{alg:lcpsimulation}
\begin{algorithmic}
\Procedure{main\_function}{$h$}
\State Initialize a global bit array $ D[2, 2n] $ to zero \Comment{$ D[2, 2n] $ marks the entries already considered}
\State \Return \Call{lcp}{$ h $}
\EndProcedure

\\

\Procedure{lcp}{$h$}
\State $ D[h] \gets 1 $
\If{$ C[h] = 1 $} \Comment{The desired value has been sampled}
    \State \Return $ \LCP^*_\mathcal{A}[rank(C, h)] $
\ElsIf{$ h $ is odd}
    \State $ i \gets \ceil{h / 2} $
    \If{$ \lambda (u_{i - 1}) \not = \lambda (u_i)$}
        \State \Return 0
    \Else{}
        \State $ k \gets p_{\max} (i - 1) $
        \State $ k' \gets p_{\min} (i) $
        \State $ j \gets RMQ_{\LCP_\mathcal{A}}(2k + 1, 2k' - 1) $
        \If{$ D[j] = 1 $} \Comment{We have already considered this entry before, so there is a cycle}
            \State \Return $ \infty $
        \Else{}
            \State \Return $ 1 + \Call{lcp}{ j } $
        \EndIf
    \EndIf    
\Else{}
    \State $ i \gets h / 2 $
    \State $ k \gets p_{\min} (i) $
    \State $ k' \gets p_{\max} (i) $
    \State $ j \gets RMQ_{\LCP_\mathcal{A}}(2k, 2k') $
    \If{$ D[j] = 1 $} \Comment{We have already considered this entry before, so there is a cycle}
        \State \Return $ \infty $
    \Else{}
        \State \Return $ 1 + \Call{lcp}{j} $
    \EndIf    
\EndIf
\EndProcedure

\\

\end{algorithmic}
\end{algorithm}

\subsubsection*{Sampling} Let $ G = (V, H) $ be a finite (unlabeled) directed graph such that every node has at most one incoming edge. For every $ v \in V $ and for every $ i \ge 0 $, there exists at most one node $ v' \in V $ such that there exists a directed path from $ v' $ to $ v $ having $ i $ edges; if $ v' $ exists, we denote it by $ v(i) $. Fix a parameter $ h \ge 1 $. Let us prove that there exists $ V(h) \subseteq V $ such that (i) $ |V(h)| \le \frac{|V|}{h} $ and (ii) for every $ v \in V $ there exists $ 0 \le i \le 2h - 2 $ such that $ v(i) $ is defined and either $ v(i) \in V(h) $ or $ v(i) $ has no incoming edges or $ v(i) = v(j) $ for some $ 0 \le j < i $. We build $ V(h) $ incrementally following Algorithm \ref{alg:V(h)construction}. Let us prove that, at the end of the algorithm, properties (i) and (ii) are true. For every $ v \in V(h) $, define $ S_v = \{v, v(1), v(2) \dots, v (h - 1) \} $, which is possible because by construction if $ v \in V(h) $, then $ v(i) $ is defined for every $ 0 \le i \le h - 1 $. It must be $ v(i) \not = v(j) $ for $ 0 \le i < j \le h - 1 $, so $ |S_v| = h $. If $ v, v' \in V(h) $ and $ v \not = v' $, then by construction $ S_v $ and $ S_{v'} $ are disjoint. As a consequence, $ |V| \ge \sum_{v \in V(h)} |S_v| = \sum_{v \in V(h)} h = h |V_h| $ and so $ |V_h| \le \frac{|V|}{h} $, which proves property (i). Let us prove property (ii). Pick $ v \in V $; we must prove that there exists $ 0 \le i \le 2h - 2 $ such that $ v(i) $ is defined and either $ v(i) \in V(h) $ or $ v(i) $ has no incoming edges or $ v(i) = v(j) $ for some $ 0 \le j < i $. We distinguish three cases:
\begin{enumerate}
    \item there exists $i$ with $ 1 \le i \le h - 1 $ such that $ v(i - 1) $ is defined but $ v(i) $ is not defined. Then, $ v(i - 1) $ has no incoming edges.
    \item there exist $i,j$ with $ 0 \le j < i \le h - 1 $ such that $ v(j) $ and $ v(i) $ are defined and $ v(i) = v(j) $. In this case, the conclusion is immediate.
    \item $ v(i) $ is defined for every $ 0 \le i \le h $ and $ v(i) \not = v(j) $ for $ 0 \le j < i \le h - 1 $. Since Algorithm \ref{alg:V(h)construction}  has terminated, then there exists $ 0 \le j \le h - 1 $ such that $ v(j) \in U $. The construction of $ U $ implies that there exists $ v' \in V $ and $ 0 \le j \le h - 1 $ such that $ v(j) = v'(j') $ and $ v'(h - 1) \in V(h) $. As a consequence $ v(h - 1 + j - j') = v(j)(h - 1 - j') = (v'(j'))(h - 1 - j') = v'(h - 1) \in V(h) $. Since $ j \le h - 1 $ and $ j' \ge 0 $, we conclude $ h - 1 + j - j' \le 2h - 2 $ and we are done.
\end{enumerate}

\begin{figure}
     \centering
     \begin{subfigure}[b]{0.59\textwidth}
         \centering
                          \resizebox{7.5 cm}{!}{
         	\begin{tikzpicture}[shorten >=1pt,node distance=1.6cm,on grid,auto]
	\tikzstyle{every state}=[fill={rgb:black,1;white,10}]

    \node[state, initial] (1) { $ u_1 $};
    \node[state] (3) [above right of=1]{ $ u_3 $};
    \node[state] (4) [below right of=1]{ $ u_4 $};    
    \node[state] (2) [above of=3]{ $ u_2 $};
    \node[state] (5) [below of=4]{ $ u_4 $}; 
    \node[state] (6) [right of=3]{ $ u_6 $}; 
    \node[state] (7) [right of=4]{ $ u_7 $};  
    \node[state, accepting] (9) [right of=6]{ $ u_9 $}; 
    \node[state, accepting] (10) [right of=7]{ $ u_{10} $};
    \node[state] (8) [above of=9]{ $ u_8 $};
    \node[state, accepting] (11) [below of=10]{ $ u_{11} $};   
    \node[state] (12) [right of=8]{ $ u_{12} $};
    \node[state] (13) [right of=9]{ $ u_{13} $};
    \node[state] (14) [right of=10]{ $ u_{14} $};
    \node[state] (15) [right of=11]{ $ u_{15} $};
    \node[state, accepting] (16) [right of=15]{ $ u_{16} $};
    
	\path[->]
    (1) edge [bend left = 50] node {a} (2) 
    (1) edge  node {b} (3)    
    (1) edge  node {c} (4) 
    (1) edge [bend right = 50] node {d} (5)  
    (2) edge [loop above] node {a}   (2)	
    (5) edge [loop below] node {d}   (5)	
    (3) edge node {e}   (6)
    (2) edge node {e}   (6)
    (4) edge node {e}   (7)
    (5) edge node {e}   (7)
    (6) edge node {f}   (8)
    (7) edge [bend right = 20] node[right] {f}   (9)
    (6) edge [bend right = 20] node[left] {g}   (10)
    (7) edge node {g}   (11)
    (8) edge node {h}   (12)	
    (12) edge node {i}   (13)	
    (13) edge node {j}   (14)	
    (14) edge node {k}   (15)	
    (15) edge node {l}   (16)	    
	;
	\end{tikzpicture}
 }
	\caption{}
     \end{subfigure}
     \hfill
     \begin{subfigure}[b]{0.40\textwidth}
         \centering
                 \resizebox{2.5 cm}{!}{
        \begin{tabular}{|c|c|c|c|c|c|c|c|}
        \hline
            State & $ i $ &  $ \LCP_\mathcal{A}[i] $ & $ k $ & $ k' $ & $ \mathcal{R}(i) $\\
        \hline
            \multirow{2}{*}{1} & 1  & & & &   \\
            & 2  & $ \infty $   & 1  & 1  & 2   \\
            \multirow{2}{*}{2} & 3 & 0 & - & - & -  \\
            & 4 & $ 1 $  & 1 & 2 & 3   \\
            \multirow{2}{*}{3} & 5 & 0  & - & - & -   \\
            & 6 & $ \infty $   & 1  & 1 & 2  \\
            \multirow{2}{*}{4} & 7 & 0  & - & - & -   \\
            & 8 & $ \infty $  & 1 & 1 & 2  \\
            \multirow{2}{*}{5} & 9 & 0  & - & - & -  \\
            & 10  & 1   & 1 & 5 & 9  \\
            \multirow{2}{*}{6} & 11 & 0 & - & - & -   \\
            & 12 & 1  & 2 & 3 & 5  \\
            \multirow{2}{*}{7} & 13 & 1 & 3 & 4 & 7  \\
            & 14 & 1  & 4 & 5 & 9  \\
            \multirow{2}{*}{8} & 15 & 0 & - & - & -  \\
            & 16 & 2  & 6 & 6 & 12  \\
            \multirow{2}{*}{9} & 17 & 2 & 6 & 7  & 13   \\
            & 18 & 2  & 7 & 7 & 14  \\            
            \multirow{2}{*}{10} & 19 & 0 & - & - & -   \\
            & 20 & 2  & 6 & 6 & 12  \\     
            \multirow{2}{*}{11} & 21 & 2 & 6 & 7 & 13   \\
            & 22 & 2  & 7 & 7 & 14  \\                 
            \multirow{2}{*}{12} & 23 & 0 & - & - & -   \\
            & 24 & 3 & 8 & 8 & 16  \\    
            \multirow{2}{*}{13} & 25 & 0 & - & - & -  \\
            & 26 & 4 & 12 & 12 & 24    \\          
            \multirow{2}{*}{14} & 27 & 0 & - & - & -  \\
            & 28 & 5 & 13 & 13 & 26  \\           
            \multirow{2}{*}{15} & 29 & 0 & - & - & -  \\  
            & 30 & 6 & 14 & 14 & 28 \\   
            \multirow{2}{*}{16} & 31 & 0 & - & - & -  \\  
            & 32 & 7 & 15 & 15 & 30 \\   
        \hline
        \end{tabular}
        }
         \caption{}
     \end{subfigure}
     \hfill
         \begin{subfigure}[b]{0.69\textwidth}
                              \resizebox{8.5 cm}{!}{     
	\begin{tikzpicture}[shorten >=1pt,node distance=1.6cm,on grid,auto]
	
    \tikzstyle{every state}=[fill={rgb:black,1;white,10}]

    \node[state] (5) { $ v_5 $};
    \node[state] (12) [right of=5]{ $ v_{12} $};z
    \node[state] (16) [above right of=12]{ $ v_{16} $};
    \node[state] (20) [below right of=12]{ $ v_{20} $};    
    \node[state, fill=yellow] (24) [right of=16]{ $ v_{24} $};
    \node[state] (26) [right of=20]{ $ v_{26} $};    
    \node[state] (28) [right of=26]{ $ v_{28} $};  
    \node[state] (30) [right of=24]{ $ v_{30} $};
    \node[state, fill=yellow] (32) [above of=30]{ $ v_{32} $};     
    \node[state] (7) [below of=5]{ $ v_7 $};    
    \node[state] (13) [below of=7]{ $ v_{13} $};
    \node[state] (17) [below left of=13]{ $ v_{17} $};
    \node[state] (21) [below right of=13]{ $ v_{21} $};
    \node[state] (9) [above right of=21]{ $ v_{9} $};
    \node[state] (10) [above right of=9]{ $ v_{10} $};
    \node[state] (14) [below right of=9]{ $ v_{14} $};
    \node[state] (18) [above right of=14]{ $ v_{18} $}; 
    \node[state] (22) [below right of=14]{ $ v_{22} $};
    \node[state] (2) [below right of=18]{ $ v_{2} $};    
    \node[state] (8) [below right of=2]{ $ v_{8} $};  
    \node[state] (6) [above right of=2]{ $ v_{6} $};  
    \node[state] (3) [above of=16]{ $ v_{3} $}; 
    \node[state] (4) [right of=3]{ $ v_{4} $}; 
    \node[state] (11) [right of=30]{ $ v_{11} $}; 
    \node[state] (15) [below of=11]{ $ v_{15} $}; 
    \node[state] (19) [below of=15]{ $ v_{19} $}; 
    \node[state] (23) [right of=11]{ $ v_{23} $}; 
    \node[state] (25) [below of=23]{ $ v_{25} $}; 
    \node[state] (27) [below of=25]{ $ v_{27} $}; 
    \node[state] (29) [below of=27]{ $ v_{29} $}; 
    \node[state] (31) [below of=29]{ $ v_{31} $};     
    
	\path[->] 
    (5) edge node {} (12)
    (12) edge node {} (16)
    (12) edge node {} (20)
    (16) edge node {} (24)
    (24) edge node {} (26)
    (26) edge node {} (28)
    (28) edge node {} (30)
    (30) edge node {} (32)    
    (7) edge node {} (13)   
    (13) edge node {} (17)  
    (13) edge node {} (21) 
    (9) edge node {} (10)  
    (9) edge node {} (14)  
    (14) edge node {} (18)  
    (14) edge node {} (22)  
    (2) edge [loop above] node {} (2) 
    (2) edge node {} (6) 
    (2) edge node {} (8)
    (3) edge node {} (4)     
	;
    
	\end{tikzpicture}
    }
	\caption{}
     \end{subfigure}
     \hfill
     \begin{subfigure}[b]{0.30\textwidth}
         \centering
        \resizebox{1.4 cm}{!}{
        \begin{tabular}{|c|c|c|}
        \hline
            $ i $ &  $ C[i] $ & $ \LCP^*_\mathcal{A} $ \\
        \hline
            1 & & 3   \\
            2 & 0 & 7     \\
            3 & 0 &   \\
            4 & 0 &     \\
            5 & 0  &     \\
            6 & 0 &    \\
            7 & 0  &    \\
            8 & 0 &   \\
            9 & 0  &    \\
            10 & 0   &    \\
            11 & 0 &    \\
            12 & 0  &   \\
            13 & 0 &    \\
            14 & 0  &    \\
            15 & 0 &   \\
            16 & 0  &    \\
            17 & 0 &     \\
            18 & 0  & \\            
            19 & 0 &     \\
            20 & 0  &    \\     
            21 & 0 &     \\
            22 & 0  &    \\                 
            23 & 0 &    \\
            24 & 1 &   \\    
            25 & 0 &    \\
            26 & 0 &      \\          
            27 & 0 &   \\
            28 & 0 &    \\           
            29 & 0 &   \\  
            30 & 0 & \\
            31 & 0 & \\
            32 & 1 & \\   
        \hline
        \end{tabular}
        }
    \caption{}
     \end{subfigure} 
        \caption{(a) A Wheeler DFA. States are numbered according to the Wheeler order. (b) The array $ \LCP_\mathcal{A} $, and the values needed to compute $ G = (V, H) $. We assume that a range minimum query returns the \emph{largest} position of a minimum value. (c) The graph $ G = (V, H) $, with $ V(\ceil{\log n}) = V(4) = \{v_{24}, v_{32} \} $ (yellow states). (d) The data structures that we store.}
        \label{fig:examplewheeler}
\end{figure}
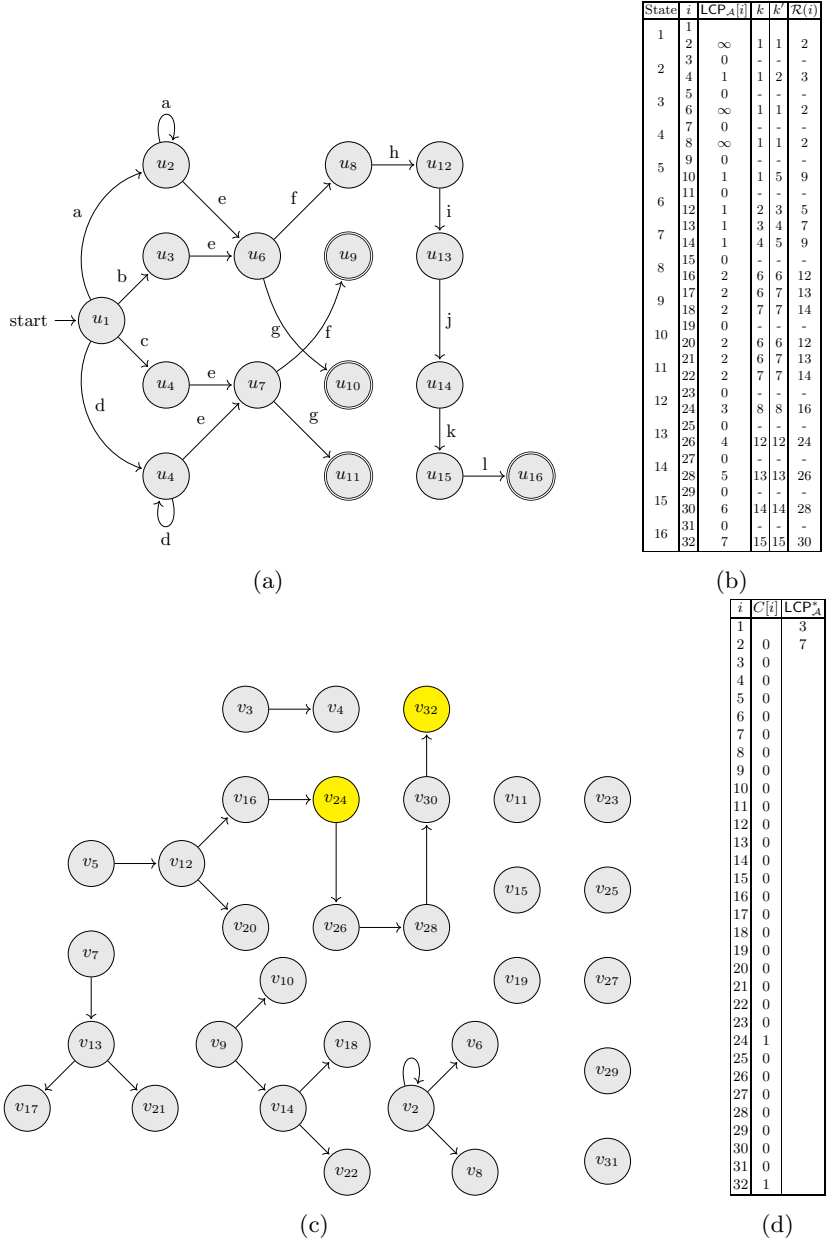

\subsubsection*{Computing the LCP Array Using a Linear Number of Bits}


First, let us store a data structure of $ O(n) $ bits which in $ O(1) $ time determines $ RMQ_{\LCP_\mathcal{A}}(i, j) $ for every $ 2 \le i \le j \le 2n $.

Notice that $ \LCP_\mathcal{A} [2i] \ge 1 $ for $ 1 \le i \le n $ because the first character of $ \min_i $ and the first character of $ \max_i $ are equal to $ \lambda (u_i) $. Moreover, we have $ \LCP_\mathcal{A} [2i - 1] \ge 1 $ if and only if $ \lambda (u_{i - 1}) = \lambda (u_i) $, for $ 2 \le i \le n $.

Consider the entry $ \LCP_\mathcal{A}[2i - 1] = \lcp(\max_{i - 1}, \min_i) $, for $ 2 \le i \le n $, and assume that $ \LCP_\mathcal{A} [2i - 1] \ge 1 $. Let $ k = p_{\max}(i - 1) $ and $ k' = p_{\min}(i) $. Since $ \LCP_\mathcal{A}[2i - 1] \ge 1 $, then there exists $ a \in \Sigma $ such that $ \max_{i - 1} = a \max_k $ and $ \min_{i - 1} = a \min_{k'} $. In particular, $ (u_{k}, u_{i - 1}, a) \in E $ and $ (u_{k'}, u_{i}, a) \in E $, so from Axiom 2 we obtain $ k < k' $. Moreover, we have $ \LCP_\mathcal{A} [2i - 1] = \lcp(\max_{i - 1}, \min_i) = \lcp(a \max_k, a \min_{k'}) = 1 + \lcp(\max_k, \min_{k'}) $. Notice that:
\begin{equation*}
\begin{split}
    & \lcp(\mathrm{max}_k, \mathrm{min}_{k'}) = \mathrm{min} \{\lcp(\mathrm{max}_k, \mathrm{min}_{k + 1}), \lcp (\mathrm{min}_{k + 1}, \mathrm{max}_{k + 1}), \dots, \\
    & = \lcp(\mathrm{min}_{k' - 1}, \mathrm{max}_{k' - 1}), \lcp(\mathrm{max}_{k' - 1}, \mathrm{min}_{k'}) \} = \\
    & = \min\{\LCP_\mathcal{A}[2k + 1], \LCP_\mathcal{A}[2k + 2], \dots, \LCP_\mathcal{A}[2k' - 2], \LCP_\mathcal{A}[2k' - 1]\}.
\end{split}
\end{equation*}

Let $ j = RMQ_{\LCP_\mathcal{A}}(2k + 1, 2k' - 1) $. Then, $ \LCP_\mathcal{A}[j] = \min\{\LCP_\mathcal{A}[2k + 1], \LCP_\mathcal{A}[2k + 2], \dots, \LCP_\mathcal{A}[2k' - 2], \LCP_\mathcal{A}[2k' - 1]\} $, so $ \LCP_\mathcal{A} [2i - 1] = 1 + \LCP_\mathcal{A} [j] $ (we assume $ t + \infty = \infty $ for every $ t \ge 0 $), and we have reduced the problem of computing $ \LCP_\mathcal{A} [2i - 1] $ to the problem of computing $ \LCP_\mathcal{A}[j] $. In the following, let $ \mathcal{R}(2i - 1) = j $. Given $ 2 \le i \le n $, we can compute $ j = \mathcal{R}(2i - 1) $ in $ O (\log \log \sigma) $ time, because we can compute $ k = p_{\max}(i - 1) $ and $ k' = p_{\min}(i) $ in $ O(\log \log \sigma) $ time and we can compute $ j $ in $ O(1) $ time by means of a range minimum query.

We proceed analogously with the entries $ \LCP_\mathcal{A} [2i] = \lcp (\min_i, \max_i) $, for $ 1 \le i \le n $ (it must necessarily be $ \LCP_\mathcal{A}[2i] \ge 1 $). Let $ k = p_{\min}(i) $ and $ k' = p_{\max}(i) $; by the definitions of $ p_{\min} $ and $ p_{\max} $ it must be $ k \le k' $. Hence, $ \LCP_\mathcal{A} [2i] = 1 + \lcp(\min_k, \max_{k'}) $ and similarly $ \lcp(\min_k, \max_{k'}) = \min\{\LCP_\mathcal{A}[2k], \LCP_\mathcal{A}[2k + 1], \dots, \LCP_\mathcal{A}[2k' ~- 1], \LCP_\mathcal{A}[2k'] \} $. Let $ j = RMQ_{\LCP_\mathcal{A}}(2k, 2k') $. In the following, let $ \mathcal{R}(2i) = j $. Given $ 1 \le i \le n $, we can compute $ j =  \mathcal{R}(2i) $ in $ O (\log \log \sigma) $ time. See Figure \ref{fig:examplewheeler} for an example.

Now, consider the (unlabeled) directed graph $ G = (V, H)  $ defined as follows. Let $ V $ be a set of $ 2n - 1 $ nodes $ v_2 $, $ v_3 $, $ \dots $, $ v_{2n} $.
Moreover, $ v_i \in V $ has no incoming edge in $ G $ if $ \mathcal{R}(i) $ is not defined, which happens if $ \LCP_\mathcal{A}[i] = 0 $ (and so $ i $ is odd and $ \lambda (u_{i - 1}) \not = \lambda (u_i) $); $ v_i \in V $ has exactly one incoming edge if $ \mathcal{R}(i) $ is defined, namely, $ (v_{\mathcal{R}(i)}, v_i) $. Note that $ v_{2i} $ has an incoming edge for every $ 1 \le i \le n $.
Let $ h \ge 1 $ be a parameter. We know that there exists $ V(h) \subseteq V $ such that (i) $ |V(h)| \le \frac{|V|}{h} $ and (ii) for every $ v_i \in V $ there exists $ 0 \le k \le 2h - 2 $ such that $ v_i (k) $ is defined and either $ v_i (h) \in V(h) $ or $ v_i (h) $ has no incoming edges or $ v_i(h) = v_i(l) $ for some $ 0 \le l < h $. Notice that if $ v_i(h) = v_i(l) $ for some $ 0 \le l < h $, then $ \LCP_\mathcal{A}[i] = \infty $ (because there is a cycle and so $ v_i(h') $ is defined for every $ h' \ge 0 $).
Let $ n' = |V(h)| $, and let $ \LCP^*_\mathcal{A}[1, n'] $ an array storing the value $ \LCP_\mathcal{A}[i] $ for each $ v_i \in V(h) $, sorted by increasing $ i $. Since $ n' \le \frac{|V|}{h} = \frac{2n - 1}{h} $, storing $ \LCP^*_\mathcal{A}[1, n'] $ takes  $ n' O(\log n) = O (\frac{n \log n}{h}) $ bits. 
We store a bitvector $ C[2, 2n] $ such that $ C[i] = 1 $ if and only if $ v_i \in V(h) $ for every $ 2 \le i \le 2n $; we augment $ C $ with $ o(n) $ bits so that it supports rank queries in $ O(1) $ time. For every $ 2 \le i \le 2n $, in $ O(1) $ time we can check whether $ \LCP_\mathcal{A}[i] $ has been stored in $ \LCP^*_\mathcal{A} $ by checking whether $ C[i] = 1 $, and if $ C[i] = 1 $ it must be $ \LCP_\mathcal{A}[i] = \LCP^*_\mathcal{A}[rank(C, i)] $.

From our discussion, it follows that Algorithm \ref{alg:lcpsimulation} correctly computes $ \LCP_\mathcal{A}[i] $ for every $ 2 \le i \le n $. Property (ii) ensures that the function $ lcp $ is called at most $ h $ times. Every call requires $ O(\log \log \sigma) $ time, so the running time of our algorithm is $ O(h \log \log \sigma) $ (the initialization of $ D[2, 2n] $ in Algorithm \ref{alg:lcpsimulation} can be simulated in $ O(1) $ time \cite{navarro2014}).
We conclude that we store $ O(n + \frac{n \log n}{h}) $ bits, and in $ O(h \log \log \sigma) $ time we can compute $ \LCP_\mathcal{A}[i] $ for every $ 2 \le i \le n $.

By choosing $ h = \ceil{\frac{\log n}{\log \log \sigma }} $, we conclude that our data structure can be stored using $ O(n \log \log \sigma) $ bits and it allows to compute $ \LCP_\mathcal{A}[i] $ for every $ 2 \le i \le n $ in $ O(\log n) $ time. By choosing $ h = \ceil{\log n} $ we conclude that our data structure can be stored using $ O(n) $ bits and it allows to compute $ \LCP_\mathcal{A}[i] $ for every $ 2 \le i \le n $ in $ O(\log n \log \log \sigma) $ time.



\section{Applications}

\subsubsection*{Matching Statistics}
Let us recall how the bounds in Theorem \ref{theor:oldmatching} are obtained. The space bound is $ O(n \log n) $ bits because we need to store $ \LCP_\mathcal{A} $. We also store a data structure to solve range minimum queries on $ \LCP_\mathcal{A} $, which only takes $ O(n) $ bits. The time bound $ O(m \log n) $ follows from performing $ O(m) $ steps to compute all matching statistics.
In each of these $ O(m) $ steps, we may need to perform a binary search on $ \LCP_\mathcal{A} $. In each step of the binary search, we need to solve a range minimum query once and we need to access $ \LCP_\mathcal{A} $ once, so the binary search takes $ O(\log n ) $ time per step. By Theorem \ref{theor:timespacetradeoff}, if we store only $ O(n \log \log \sigma) $ bits, we can access $ \LCP_\mathcal{A} $ in $ O(\log n) $ time, so the time for the binary search becomes $ O(\log^2 n) $ per step and Theorem \ref{theor:newmatching} follows.

\subsubsection*{Variable-order de Bruijn Graphs}

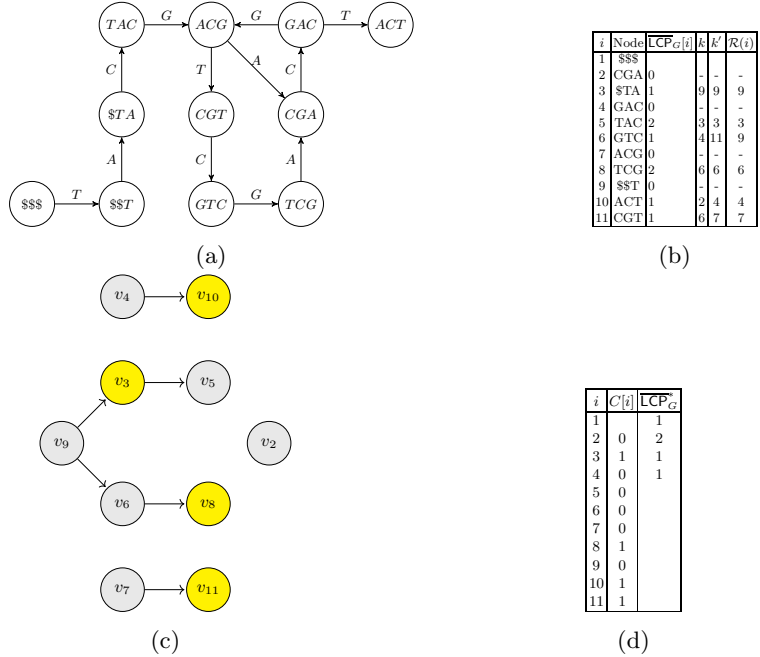
\begin{figure}[!t]
     \centering
     \begin{subfigure}[b]{0.59\textwidth}
         \centering
                                   \resizebox{5.5 cm}{!}{
\begin{tikzpicture}[->,>=stealth', semithick, initial text={}, auto, scale=.42]
 \node[state, minimum size=30pt,] (1) at (0,0) {$\$\$\$$};
 \node[state, label=above:{}, minimum size=30pt] (2) at (5,0) {$\$\$T$};
  \node[state, label=above:{}, minimum size=30pt] (3) at (5,5) {$\$TA$};
 \node[state, label=above:{}, minimum size=30pt] (4) at (5,10) {$TAC$};
  \node[state, label=above:{}, minimum size=30pt] (5) at (10,10) {$ACG$}; 
 \node[state, label=above:{}, minimum size=30pt] (6) at (10,5) {$CGT$};
 \node[state, label=above:{}, minimum size=30pt] (7) at (10,0) {$GTC$};
  \node[state, label=above:{}, minimum size=30pt] (8) at (15,0) {$TCG$};
  \node[state, label=above:{}, minimum size=30pt] (9) at (15,5) {$CGA$};
  \node[state, label=above:{}, minimum size=30pt] (10) at (15,10) {$GAC$};
 \node[state, label=above:{}, minimum size=30pt] (11) at (20,10) {$ACT$}; 
 \draw (1) edge [above] node [above] {$T$} (2);
  \draw (2) edge [above] node [left] {$A$} (3);
   \draw (3) edge [above] node [left] {$C$} (4);
    \draw (4) edge [above] node [above] {$G$} (5);
     \draw (5) edge [above] node [left] {$T$} (6);
      \draw (6) edge [above] node [left] {$C$} (7);
       \draw (7) edge [above] node [above] {$G$} (8);
        \draw (8) edge [above] node [left] {$A$} (9);
         \draw (9) edge [above] node [left] {$C$} (10);
          \draw (10) edge [above] node [above] {$T$} (11);
          \draw (5) edge [above] node [above] {$A$} (9);
          \draw (10) edge [above] node [above] {$G$} (5);
\end{tikzpicture}
}
	\caption{}
     \end{subfigure}
     \hfill
     \begin{subfigure}[b]{0.40\textwidth}
         \centering
        \centering
                         \resizebox{2.3 cm}{!}{
        \begin{tabular}{|c|c|l|c|c|c|c|c|}
        \hline
        & & & & & \\[-1em]
            $ i $ & Node & $ \overline{\LCP}_G[i] $ & $ k $ & $ k' $ & $ 
            \mathcal{R}(i) $\\
        \hline
            1 & \$\$\$ & & & &   \\
            2 & CGA & 0 & - & - & -  \\
            3 & \$TA & 1 & 9 & 9 & 9 \\
            4 & GAC & 0 & - & - & -  \\
            5 & TAC & 2 & 3 & 3 & 3 \\
            6 & GTC & 1 & 4 & 11 & 9 \\
            7 & ACG & 0 & - & - & -  \\
            8 & TCG & 2 & 6 & 6 & 6 \\
            9 & \$\$T & 0 & - & - & - \\
            10 & ACT & 1 & 2 & 4 & 4 \\
            11 & CGT & 1 & 6 & 7 & 7 \\
        \hline
        \end{tabular}
        }
    \caption{}
     \end{subfigure}
     \hfill
         \begin{subfigure}[b]{0.49\textwidth}
		\centering         
         \resizebox{3.5 cm}{!}{ 
         \centering
	\begin{tikzpicture}[shorten >=1pt,node distance=1.6cm,on grid,auto]
	
    \tikzstyle{every state}=[fill={rgb:black,1;white,10}]

    \node[state] (9) { $ v_9 $};
    \node[state, fill = yellow] (3) [above right of=9]{ $ v_{3} $};  
    \node[state] (6) [below right of=9]{ $ v_{6} $};  
    \node[state] (5) [right of=3]{ $ v_{5} $};  
    \node[state, fill = yellow] (8) [right of=6]{ $ v_{8} $};  
    \node[state] (4) [above of=3]{ $ v_{4} $};  
    \node[state, fill = yellow] (10) [right of=4]{ $ v_{10} $};  
    \node[state] (7) [below of=6]{ $ v_{7} $};  
    \node[state, fill = yellow] (11) [right of=7]{ $ v_{11} $}; 
    \node[state] (2) [below right of=5]{ $ v_{2} $}; 
    
	\path[->]     
    (9) edge node {} (3)
    (9) edge node {} (6)
    (3) edge node {} (5)
    (6) edge node {} (8)
    (4) edge node {} (10)
    (7) edge node {} (11)
	;
    
	\end{tikzpicture}
	}
	\caption{}
     \end{subfigure}
     \hfill
     \begin{subfigure}[b]{0.49\textwidth}
         \centering
        \resizebox{1.4 cm}{!}{
        \begin{tabular}{|c|c|c|}
        \hline
        & &  \\[-1em]
            $ i $ &  $ C[i] $ & $ \overline{\LCP}^*_G $ \\
        \hline
            1 & & 1   \\
            2 & 0 & 2     \\
            3 & 1 & 1  \\
            4 & 0 & 1   \\
            5 & 0  &     \\
            6 & 0 &    \\
            7 & 0  &    \\
            8 & 1 &   \\
            9 & 0  &    \\
            10 & 1   &    \\
            11 & 1 &    \\
        \hline
        \end{tabular}
        }
    \caption{}
     \end{subfigure} 
        \caption{The $ 3 $-rd order de Bruijn graph for the set $ \mathcal{S} = \{CGAC,
GACG, GACT, TACG, GTCG, ACGA, ACGT, TCGA, CGTC \}$ from \cite{boucher2015}. 
We proceed like in Figure \ref{fig:examplewheeler} (now we only consider odd entries of $ \LCP_G $, and $ h = \ceil{\log k} = 2 $).}
\label{fig:exampledebruijn}
\end{figure}

Let $ k \ge 0 $ be a parameter, and let $ \mathcal{S} $ be a set of strings on the alphabet $ \Sigma = \{A, C, G, T \} $ (in this application we always assume $ \sigma = O(1) $). 

The $ k $-th order de Bruijn graph of $ \mathcal{S} $ is defined as follows. The set of nodes is the set of all strings of $ \Sigma $ of length $ k $ that occur as a substring of some string in $ \mathcal{S} $. There is an edge from node $ \alpha $ to node $ \beta $ labeled $ c \in \Sigma $ if and only if (i) the suffix of $ \alpha $ of length $ k - 1 $ is equal to the prefix of $ \beta $ of length $ k - 1 $ and (ii) the last character of $ \beta $ is $ c $. If some node $ \alpha $ has no incoming edges, then we add nodes $ \$^i \alpha_{k - i} $ for $ 1 \le i \le k $, where $ \alpha_j $ is the prefix of $ \alpha $ of length $ j $ and $ \$ $ is a special character, and we add edges as above; see Figure \ref{fig:exampledebruijn} for an example. 
Wheeler DFAs are a generalization of de Bruijn graphs (we do not need to define an initial state and a set of final states, because here we are not interested in studying the applications of de Bruijn graphs and Wheeler automata to automata theory \cite{alanko2021, cotumaccio2023}); the Wheeler order is the one such that node $ \alpha $ comes before node $ \beta $ if and only if the string $ \alpha^R $ is lexicographically smaller than the string $ \beta^R $ \cite{gagie2017}.

Notice that, in a $ k $-th order de Bruijn graph $ G $, all strings that can be read from node $ \alpha $ by following edges in a backward fashion start with $ \alpha^R $ (as usual, we assume that node $ \$\$\$ $ has a self-loop labeled $ \$ $). As a consequence, it holds $ \LCP_G[2i] \ge k $ for every $ 1 \le i \le n $ and $ \LCP_G[2i - 1] \le k - 1 $ for every $ 2 \le i \le n $ (so any value in an odd entry is smaller than any value in an even entry).

As we saw in the introduction, in \cite{boucher2015} it was shown that the $ k $-order de Bruijn graph of $ \mathcal{S} $ can be used to \emph{implicitly} store the $ k' $-th order de Bruijn graph of $ \mathcal{S} $ for \emph{every} $ k' \le k $, thus leading to a \emph{variable-order de Bruijn graph}. The navigation of a variable-order de Bruijn graph is possible by storing or by simulating the values in the odd entries of the LCP array. Formally, in order to avoid confusion, we define $ \overline{\LCP}_G[i] = \LCP_G[2i - 1] $ for every $ 2 \le i \le n $; see Figure \ref{fig:exampledebruijn}. Note that $ \overline{\LCP}_G[i] \le k - 1 $ for every $ 2 \le i \le n $, so $ \overline{\LCP}_G $ can be stored by using $ O(n \log k) $ bits. Notice that Theorem \ref{theor:timespacetradeoff} also applies to $ \overline{\LCP}_G[i] $ (we do not need to store values in the even entries because a value in an odd entry is smaller than a value in an even entry, so even entries are never selected in the sampling process when answering a range minimum query on $ \LCP_G $). However, we can now choose a better parameter $ h \ge 1 $ in our sampling process. Indeed, each entry of $ \overline{\LCP}_G $ can be stored by using $ O(\log k) $ bits (not $ O(\log n $) bits), so if we choose $ h = \ceil{\log k} $, we conclude that we can augment the BOSS representation of a de Bruijn graph with $ O(n ) $ bits such that for every $ 2 \le i \le n $ we can compute $ \overline{\LCP}_G[i] $ in $ O(\log k ) $ time.

The first solution in Theorem \ref{theor:oldvaribale} consists in storing a wavelet tree on $ \overline{\LCP}_G $, which requires $ O(n \log k ) $ bits and allows to navigate the graph in $ O( \log k) $ time per visited node. 
The second solution in Theorem \ref{theor:oldvaribale} does not store $ \overline{\LCP}_G $ at all; whenever needed, an entry of $ \overline{\LCP}_G $ is computed in $ O(k) $ time by exploiting the BOSS representation of the de Bruijn graph. The second solution only stores a data structures of $ O(n) $ bits to solve range minimum queries. The details can be found in \cite{boucher2015}. Essentially, the time bound $ O(k \log n) $ comes from performing binary searches on  $ \overline{\LCP}_G $ while explicitly computing an entry of $ \overline{\LCP}_G $ at each step in $ O(k) $ time. However, we have seen that, while staying within the $ O(n) $ space bound, we can augment the BOSS representation so that we can compute the entries of $ \overline{\LCP}_G $ in $ O(\log k) $ time, so the time bound $ O(k \log n) $ becomes $ O(\log k \log n) $, which implies Theorem \ref{theor:newvariable}.

%
%
%

\subsubsection*{Acknowledgements}

\emph{Travis Gagie}: funded by National Institutes of Health (NIH) NIAID (grant no. HG011392),
the National Science Foundation NSF IIBR (grant no. 2029552) and a Natural Science and
Engineering Research Council (NSERC) Discovery Grant (grant no. RGPIN-07185-2020). \emph{Dominik Köppl}: supported by JSPS KAKENHI with No. JP21K17701 and JP23H04378. \emph{Nicola Prezza}: funded by the European Union (ERC, REGINDEX, 101039208). Views and opinions expressed are however those of the author(s) only and do not necessarily reflect those of the European Union or the European Research Council. Neither the European Union nor the granting authority can be held responsible for them.

\bibliographystyle{splncs04}
\bibliography{mybibliography}
%






\end{document}